\documentclass[12pt]{article}
\usepackage{graphicx}
\usepackage{amssymb}
\usepackage{amsmath}
\usepackage{subfigure}
\usepackage{cite}
\input{colordvi.tex}

\begin{document}

\begin{titlepage}

\begin{flushright}
ICRR-Report-690-2014-16\\
\end{flushright}

\vskip 1.35cm

\begin{center}

{\large 
{\bf Thermal Effects and Sudden Decay Approximation \\ in the Curvaton Scenario} 
}

\vskip 1.2cm

Naoya Kitajima$^{a,b}$,
David Langlois$^c$,
Tomo Takahashi$^d$,
Tomohiro Takesako$^a$
and
Shuichiro Yokoyama$^a$ \\

\vskip 0.4cm

{ \it$^a$Institute for Cosmic Ray Research,
University of Tokyo, Kashiwa 277-8582, Japan}\\
{\it$^b$Department of Physics, Tohoku University, Sendai 980-8578, Japan}\\
{\it$^c$Astroparticle \& Cosmologie (CNRS-Universit\'e Paris 7),
10 rue Alice Domon et L\'eonie Duquet, 75205 Paris Cedex 13, France}\\
{\it $^d$Department of Physics, Saga University, Saga 840-8502, Japan }\\

\date{\today}

\begin{abstract} 
We study the impact of a temperature-dependent curvaton decay rate on the primordial curvature  
perturbation generated in the curvaton scenario.
Using the familiar sudden decay approximation, we obtain an analytical expression for the curvature perturbation after the decay of the curvaton. 
We then  investigate numerically the evolution of the background and of the perturbations during the decay. 
We first show that the instantaneous transfer coefficient, related to the curvaton energy fraction at the decay, can be extended into a more general parameter, which depends on the net transfer of the curvaton energy into radiation energy or, equivalently, on the total entropy ratio after the complete curvaton decay.
We then compute the curvature perturbation and compare this result with  the sudden decay approximation prediction. 

\end{abstract}

\end{center}
\end{titlepage}

\section{Introduction}
\label{sec:intro}

The improving  precision  of cosmological observations  requires in parallel 
more refined analyses of the physics of the
 early Universe, especially concerning the mechanisms 
 responsible for the primordial curvature perturbations that act as seeds of  the cosmic microwave background (CMB)
anisotropies and structure formation in our Universe.
Although the simplest  scenarios assume that the primordial fluctuations originate 
directly
from the quantum fluctuations of a scalar field, 
which drives a quasi-exponential expansion of the early Universe, called inflation~\cite{Guth:1980zm, Starobinsky:1980te, Sato:1980yn, Linde:1981mu, Albrecht:1982wi}, an interesting variant is 
the curvaton scenario~\cite{Lyth:2001nq, Moroi:2001ct, Enqvist:2001zp}, where the primordial fluctuations are generated from a light scalar field other than the inflaton, the so-called curvaton. One can also envisage hybrid scenarios where the fluctuations of both the inflaton and curvaton are relevant~\cite{Langlois:2004nn,Moroi:2005kz,Moroi:2005np,Ichikawa:2008iq,Enqvist:2013paa}\footnote{
If the CMB B-mode polarization detected by BICEP2 \cite{Ade:2014xna}  corresponds to a primordial signal, this would   indicate that the energy scale of 
inflation is relatively high, $H_{\rm inf} \sim 10^{14}~{\rm GeV}$. With such high inflationary energy scale, 
fluctuations from the inflaton  also tend to contribute to the curvature perturbation
even in the curvaton model \cite{Ichikawa:2008iq,Enqvist:2013paa,Byrnes:2014xua,Fujita:2014iaa}.
Thus the model naturally becomes a hybrid (mixed) model. We should also mention another possibility where 
the inflaton is trapped at the false vacuum during inflation and has a large mass. In such a case, 
 fluctuations from the inflaton is suppressed and the curvature perturbation can be purely dominated by 
those from the curvaton \cite{Kehagias:2014wza}.
}.

In the standard curvaton scenario, the curvaton field is supposed to start oscillating during the radiation dominated phase, when it is surrounded by the thermal plasma generated by the reheating at the end of inflation. Isocurvature fluctuations of the curvaton can later be transferred into  the curvature perturbation, via the decay of the curvaton into radiation.
Since the curvaton is necessarily coupled to the thermal plasma, thermal effects of the plasma could affect not only the dynamics of the curvaton decay but also the produced primordial curvature perturbations. The latter aspects  have not been discussed much in the literature\footnote{
In Ref.~\cite{Lyth:2003ip}, the temperature dependence of the CDM annihilation rate is taken into account.
Recently, Ref. \cite{Mukaida:2014yia} has also evaluated the curvature perturbation in the case with the temperature/curvaton field-dependent decay rate
by using the sudden decay approximation.
}.

In this paper, we present an attempt to discuss these effects, by  simply taking into account the  dependence of the curvaton decay  rate $\Gamma$ on  the cosmic temperature $T$
in the evaluation of the curvature perturbation $\zeta$.
Note that such a temperature dependent decay rate $\Gamma(T)$ is quite generic and  can be computed explicitly  when the plasma is  at sufficiently high temperatures (see Refs.~\cite{Mukaida:2014yia,Yokoyama:2005dv, Parwani:1991gq, Yokoyama:2004pf, Drewes:2010pf, Drewes:2013iaa, Bodeker:2006ij, Laine:2010cq, Arnold:2006fz, Moroi:2012vu, Mukaida:2012qn}).
For instance, typical interactions like $\mathcal{L}_{\text{int}} = - M \sigma \chi^2 - \lambda \chi \xi^2$
or $\mathcal{L}_{\text{int}} = - y \sigma \bar \psi \psi - g A_{\mu} \bar \psi \gamma^{\mu} \psi$ can induce $\Gamma (T)$~\cite{Drewes:2013iaa}
where $M, \lambda, y, g$ are couplings and the temperature $T$ should be much greater than the masses of the non-thermalized curvaton field $\sigma$
and the other fields ($\chi, \xi, \psi, A_{\mu}$) in the plasma in order that the thermal effect becomes significant.

Although it is common to adopt the so-called sudden decay approximation in calculating 
the curvature perturbation in the curvaton model,
this approximation does not always provide an accurate description, in particular when one considers the thermal effects mentioned above. In the present work,
we identify in which cases the sudden decay approximation is valid or not, by comparing the analytical predictions with the numerical results. When the approximation does not give an accurate prediction, 
one needs to resort to a numerical calculation to discuss the thermal effects on the curvature perturbation.

This paper is organized as follows. 
In Sec.~\ref{sec:setup}, we briefly present the curvaton model and introduce a temperature-dependent decay rate of the curvaton $\Gamma (T)$.
In Sec.~\ref{sec:ana}, we use the sudden decay approximation to make an analytical estimate of the perturbations.
In Sec.~\ref{sec:num}, we solve numerically the evolution equations for the coupled system consisting of the curvaton and   radiation, and  compare these results with the analytical expressions obtained in Sec.~\ref{sec:ana}.
We also discuss the validity of the sudden decay approximation there.
Sec.~\ref{sec:conc} is devoted to our conclusions.

\section{Set-up}
\label{sec:setup}

In this section, we present a model that  will be investigated in the rest of this paper.
We consider the simplest curvaton scenario in which we distinguish only two components: the curvaton and radiation. We start our analysis during the phase when the curvaton is oscillating at the bottom of its potential (assumed to be quadratic). As a consequence, the curvaton can be treated as an effective fluid with vanishing pressure.
For simplicity,
we neglect the contribution to 
the curvature perturbation from the inflaton field  throughout the whole discussion in this paper, although it can be included in a straightforward way. 

In a spatially flat FLRW universe, characterized by the metric 
\begin{equation}
ds^2=-dt^2+a^2(t)d \vec{x}^2\,,
\end{equation}
the background evolution equations for the curvaton and radiation energy densities, respectively denoted as $\rho_\sigma$  and $\rho_r$, are given by
\begin{equation}\label{eq:evolvrho}
\begin{split}
\frac{\mathrm{d}}{\mathrm{d} t} \rho_{\sigma}
&= - 3 H \rho_{\sigma} - \Gamma (T) \rho_{\sigma}, \\
\frac{\mathrm{d}}{\mathrm{d} t} \rho_r
&= - 4 H \rho_r + \Gamma (T) \rho_{\sigma},\\
H^2
&= \frac{1}{3 M_{\text{P}}^2} (\rho_{\sigma} + \rho_r),
\end{split}
\end{equation}
where $H\equiv (da/dt)/a$ is the Hubble parameter, $\Gamma (T)$ is a decay rate of the curvaton
which depends on the cosmic temperature, $T$,
and $M_{\text{P}}\equiv (8\pi G)^{-1/2} \simeq 2.4 \times 10^{18}~\mathrm{GeV}$ is the reduced Planck mass. 
The above description is valid from  an initial time $t_i$ corresponding to the onset of the curvaton oscillations with $H \sim m_{\sigma}$, when  
the total energy density of the universe is dominated by that of radiation. The curvaton and  radiation
energy densities then evolve according to Eqs.~(\ref{eq:evolvrho}).  
When the Hubble parameter reaches $H \sim \Gamma (T)$, the curvaton starts to decay and its energy density  is  very rapidly transferred into that of  radiation.

As for the perturbations, the existence of initial fluctuations $\delta \sigma_i$  of a subdominant curvaton field implies the presence of the initial  isocurvature perturbation $\mathcal{S}$, which  is given by 
\begin{equation}
\begin{split}
\mathcal{S}_i
&= 3 (\zeta_{\sigma, i} - \zeta_{r, i})
\,.
\end{split}
\label{eq:iniiso}
\end{equation}
In the above expression,  the subscript $i$ indicates that the corresponding quantity is evaluated  at the initial time $t_i$.  
We have also introduced the curvature perturbation $\zeta_a$ for each component $a$, which is defined nonlinearly as \cite{Lyth:2004gb} (see also
\cite{Langlois:2005ii,Langlois:2005qp,Langlois:2008vk,Langlois:2010vx} for a covariant
definition)
\begin{equation}
\label{zeta_a}
\zeta_{a} = \delta N + \frac{1}{3(1+w_a)} \ln \left( \frac{\rho_a ( t, \vec{x})}{\bar{\rho}_a(t)} \right), 
\end{equation}
where $\delta N$ denotes the local perturbation of the number of
$e$-folds, $w_a\equiv \bar{P}_a/ \bar{\rho}_a$ is 
the equation of state for a fluid $a$, which is
assumed to be constant,   and a barred quantity must be understood as the homogeneous one.

In the following, for explicit calculations, we  assume  a specific functional form of $\Gamma (T)$ given as
\begin{equation}\label{eq:Gamma}
\begin{split}
\Gamma (T) = \Gamma_0 \left[ A + {C \left( T / m_{\sigma} \right)^n \over 1+ C \left( T / m_{\sigma} \right)^n} \right],
\end{split}
\end{equation}
where $m_{\sigma}$ is the zero-temperature mass of the curvaton, $\Gamma_0$, $A  \ll 1$ and $C$ are constant parameters.
A constant index $n$ determines the power of the temperature dependence of $\Gamma (T)$.
Although the actual temperature dependence of $\Gamma (T)$ is far more complicated than Eq.~(\ref{eq:Gamma}), 
we will use Eq.~(\ref{eq:Gamma}) as a simple model for the decay rate.
For more realistic form of $\Gamma (T)$, see for example Refs.~\cite{Drewes:2010pf, Drewes:2013iaa}.
For the above form of $\Gamma (T)$,
in the high temperature limit $T \to \infty$, the decay rate becomes as $\Gamma (T) \to \Gamma_0 \left( 1 + A \right)$.
On the other hand, in the low temperature limit $T \to 0$, one has $\Gamma (T) \to \Gamma_0 A$.
In the intermediate temperature range where $ A < C(T/m_\sigma)^n < 1 $,
the decay rate depends on the temperature as $\Gamma (T) \simeq \Gamma_0 C (T/m_\sigma)^n$.
If the decay rate can be approximated as $\Gamma (T) \sim  T^n$ at the time around when $H \sim \Gamma (T)$
as in the intermediated case mentioned above,
such temperature dependence is expected to affect the final curvature perturbations, as will be discussed later.

\section{Sudden decay approximation}
\label{sec:ana}
In this section, we attempt to derive an analytical estimate of the final curvature perturbation, by adopting
 the sudden decay approximation which has been widely used in the context of the standard curvaton scenario. In this approximation, 
 the curvaton is supposed to decay instantaneously, which  leads to the notion of a spacelike decay hypersurface, on which one can explicitly 
compute  the perturbations.

Inverting the nonlinear definition (\ref{zeta_a}) of the curvature perturbation for each individual fluid, one can express the  energy density of the species $a$ in the form
\begin{equation}
\label{rho_a}
\rho_a (t, \vec{x}) = \bar{\rho}_a(t) e^{3 (1+ w_a) (\zeta_a - \delta N)}.
\end{equation}
In our case, we need to  consider only two  species: radiation
($w_r=1/3$)
and  the curvaton field, treated as a pressureless fluid
($w_\sigma=0$). Note that this can be generalized to an arbitrary number of species~\cite{Langlois:2011zz}.

On the decay hypersurface, characterized by $H=\Gamma$ and the perturbation $\delta N_D$, one can write, just before the decay \cite{Langlois:2013dh}
\begin{equation}
\label{rho_total_before}
\rho_{\rm total}= \rho_{\sigma}+\rho_r= \bar{\rho}_\sigma\mathrm{e}^{3 (\zeta_{\sigma, i} - \delta N_D)}
+\bar{\rho}_r\mathrm{e}^{4 (\zeta_{r, i} - \delta N_D)}=3 M_{\text{P}}^2 \Gamma^2=3 M_{\text{P}}^2 \bar\Gamma^2(1+\delta_\Gamma)^2\,,
\end{equation}
where we have introduced the relative fluctuations $\delta_\Gamma$ on the decay hypersurface, defined as
\begin{equation}
\Gamma\equiv\bar\Gamma\left(1+\delta_\Gamma\right)\,.
\end{equation}
Indeed, $\Gamma$  is in general nonuniform on the decay hypersurface, since it depends on the temperature which can fluctuate. 
Expanding the above relation (\ref{rho_total_before}), one finds at linear order
\begin{equation}
\label{N_D1}
\delta N_D=\frac{1}{3\Omega_\sigma+4\Omega_r}\left(3\Omega_\sigma \zeta_{\sigma,i} +4\Omega_r\zeta_{r,i}-2\delta_\Gamma\right)\,,
\end{equation}
where the parameters $\Omega_a \equiv \bar{\rho}_a/\bar{\rho}_{\rm tot}$ denote the energy density fractions just before the decay, and satisfy $\Omega_\sigma+\Omega_r=1$.

The relation between the total energy density and the decay rate of the curvaton
at the decay hypersurface, expressed in (\ref{rho_total_before}), can also be written as 
\begin{equation}\label{eq:justafter}
\begin{split}
\bar\rho_{\text{tot}}\,\mathrm{e}^{4 (\zeta - \delta N_D)}=3 M_{\text{P}}^2 \bar{\Gamma}^2(1+\delta_\Gamma)^2\,,
\end{split}
\end{equation}
which implies the following  relation between $\zeta$,  $\delta N_D$ and  the decay rate fluctuation
\begin{equation}\label{eq:atdecay}
\begin{split}
\zeta = \delta N_D + \frac{1}{2} \delta_{\Gamma},
\end{split}
\end{equation}
at  linear order. 
Combining Eqs.~(\ref{N_D1}), (\ref{eq:atdecay}) 
and the definition (\ref{eq:iniiso}), we finally get
\begin{equation}\label{eq:zeta1st}
\begin{split}
\zeta = \zeta_{\text{inf}} + \frac{r_{\rm dec}}{3} \mathcal{S}_i - \frac{r_{\rm dec}}{6} \delta_{\Gamma},
\end{split}
\end{equation}
with
\begin{equation}
\label{r_dec}
\zeta_{\rm inf} \equiv \zeta_{r,i},~~
r_{\rm dec}\equiv \frac {3\Omega_{\sigma,{\rm dec}}}{4-\Omega_{\sigma,{\rm dec}}} 
= {3\bar{\rho}_{\sigma,{\rm dec}} \over 3 \bar{\rho}_{\sigma,{\rm dec}} + 4 \bar{\rho}_{r,{\rm dec}}}\,,
\end{equation}
where the subscript ``dec'' means that the corresponding quantities are evaluated on the decay hypersurface,  just before the decay.

The  expression for $\zeta$ given in Eq.~\eqref{eq:zeta1st}
has exactly the same form as the one obtained for the modulated decay  of the curvaton~\cite{Langlois:2013dh, Assadullahi:2013ey,Enomoto:2013qf}. However, in the latter scenario, the fluctuations 
 $\delta_\Gamma$ originate from
a light scalar field other than the inflaton or the curvaton. By contrast, in the present case, the fluctuations $\delta_\Gamma$ arise from the temperature dependence of the decay rate
\cite{Mukaida:2014yia}, and are given, at linear order, by
\begin{equation}
\label{d_Gamma}
\delta_\Gamma=\alpha\, \left(\frac{\delta T}{T}\right)_{_D}\,, \qquad \alpha\equiv \frac{d\ln \Gamma}{d\ln T}\,.
\end{equation}
Moreover, since $\rho_r\propto T^4$, one can also write
\begin{equation}
\label{d_T}
\left(\frac{\delta T}{T}\right)_{_D}=\frac14\left(\frac{\delta \rho_r}{\rho_r}\right)_{_D}= \zeta_{r,i} -\delta N_D\,,
\end{equation}
where we have used (\ref{rho_a}) for the last equality. Inserting (\ref{d_Gamma}) and (\ref{d_T}) into (\ref{N_D1}), one finally gets the expression of $\delta N_D$ in terms of the two perturbations $\zeta_r$ and $\zeta_\sigma$ defined just before the decay:
\begin{equation}
\label{N_D2}
\delta N_D=\frac{1}{3\Omega_\sigma+4\Omega_r-2\alpha}\left[3\Omega_\sigma \zeta_{\sigma,i} +\left(4\Omega_r-2\alpha\right)\zeta_{r,i}\right]\,.
\end{equation}
Therefore, the expression of  $\delta_\Gamma$ in terms of the isocurvature perturbation is 
\begin{eqnarray}
\delta_\Gamma &=& -  \alpha \, \frac{3 r_{\rm dec}}{3+r_{\rm dec}}
\left( \frac{3}{3+r_{\rm dec}}- \frac{\alpha}{2} \right)^{-1} {{\mathcal S_i} \over 3} \,.
\label{eq:dGS}
\end{eqnarray}
Inserting this result into (\ref{eq:zeta1st}) yields the prediction for the final curvature perturbation in terms of $\zeta_{\rm inf}$ and ${\mathcal S_i}$, within the sudden decay approximation.

The above expression (\ref{eq:dGS}) implies the a priori surprising result that the quantity $\delta_\Gamma$, and thus the final curvature perturbation $\zeta$,  can be strongly enhanced by a special tuning of the parameters $r_{\rm dec}$ and $\alpha$. This result can in fact be understood as the consequence of an accidental degeneracy. For simplicity, let us consider a perfectly homogeneous and isotropic Universe, without perturbation. The decay hypersurface ($N=N_D$) is defined by the condition $H=\Gamma(T)$. A small deformation $\delta N_D$ of this hypersurface leads to a (linear) variation of the total energy density,
\begin{equation}
\frac{\delta\rho_{\rm total}}{\bar\rho_{\rm total}}=\left[-3\Omega_\sigma-4\left(1-\Omega_\sigma\right)\right]\delta N_D\,,
 \end{equation}
 as well as to a variation of the decay rate squared,
\begin{equation}
 \frac{\delta (\Gamma^2)}{\bar\Gamma^2}=-2\alpha\delta N_D\,.
\end{equation}
Comparison of  the two above expressions shows that a slightly deformed hypersurface still satisfies the equality between $H$ and $\Gamma$ at the linear level, provided $4-\Omega_\sigma=2\alpha$. Therefore, when the parameters satisfy this tuning, the decay hypersurface is degenerate,  at least at the linear level. This degeneracy explains why the position $\delta N_D$ of the decay hypersurface is very sensitive to the value of cosmological perturbations when the combination $4-\Omega_\sigma-2\alpha$ is close to zero. 
 Obviously, this conclusion strongly relies on the assumption that the decay is localized on a well-defined hypersurface.  From a more realistic point of view, the notion of a decay hypersurface is blurred and the artificial enhancement predicted by the sudden decay approximation is not expected to remain valid. This will be confirmed in our numerical study presented in the next section.

In the next section, we evaluate the curvature perturbation by performing a numerical calculation, 
and compare the numerical results with the analytic one derived by adopting the sudden decay approximation in this section.

\section{Numerical results}
\label{sec:num}

In this section, we compute numerically the evolution of the background  and  of the perturbations. We first revisit the constant decay rate case by introducing a more general definition for the transfer coefficient, which  turns out to provide a very good fit to the numerics. We then consider  cases with  a temperature dependent decay rate and discuss our  numerical procedure and results. 
  Finally, we analyse in more detail the discrepancies between  the analytic formulas obtained with  the sudden decay approximation  and our numerical results.

\subsection{A new definition of the transfer parameter}

In order to compare the analytic and numerical results, one must be aware  that 
$r_{\rm dec}$ is not a very good parameter to evaluate the final curvature perturbation, even in the constant decay rate case ($\delta_\Gamma = 0$ case). In contrast with  the sudden decay approximation, where the curvaton is  instantaneously  transferred into radiation,  the curvaton starts to gradually decay into radiation before $H=\Gamma$ and its decay product then becomes part of 
radiation,  which can be followed by numerically solving  Eqs.~\eqref{eq:evolvrho}. Consequently, 
the value of $r_{\rm dec}$ evaluated in the numerical calculation is expected to be smaller than the estimate provided by the sudden decay approximation.

For the standard constant decay rate case ($n=0$), 
it has already been noticed~\cite{Malik:2002jb,Sasaki:2006kq,Gupta:2003jc} that a better agreement with the numerical result can be obtained by replacing the parameter $r_{\text{dec}}$ 
with a fitting parameter $r_{\rm fit}$ of the form 
\begin{equation}
r_{\text{fit}} = 1 - \left( 1 + \frac{\beta}{\gamma} p \right)^{-\gamma}\,, \qquad p = ( \Omega_{\sigma} \sqrt{ H / \Gamma} )|_{\bar H = m_{\sigma}}\,.
\end{equation}
The values $\beta=0.924$ and $\gamma=1.24$ provide a good fit to the numerical result. Note that one recovers $r_{\rm dec}$ by  replacing $p$ with $\bar{\rho}_{\sigma, {\rm dec}} / \bar{\rho}_{r,{\rm dec}}$
and using $\beta=3/4$ and $\gamma=1$  respectively in the above expression.
In the constant $\Gamma$ case, since $\Omega_\sigma \propto a$ and $H \propto a^{-2}$ (in a radiation dominated Universe),
the parameter $p$ is conserved before the curvaton decay and 
 thus represents an appropriate parameter to describe the curvature perturbation in the curvaton scenario.
However, in the case of a temperature-dependent decay rate,
the parameter $p$ is no longer a well-defined conserved quantity  because the decay rate $\Gamma$ varies with time.

For this reason,  we introduce a new definition of the transfer parameter, given by  
\begin{equation}
\begin{split}
\label{r_s}
r_s
&\equiv \frac{3 \bar\rho_{r \sigma, f}}{3 \bar \rho_{r \sigma, f} + 4  \bar\rho_{r \phi, f}}\,,
\end{split}
\end{equation}
where $\bar{\rho}_{r\sigma,f}$ and $\bar{\rho}_{r\phi,f}$ are
respectively  the energy densities of radiation components sourced by the curvaton and the inflaton, well after the curvaton decay. Even in the temperature dependent $\Gamma$ case, $r_s$ is well defined since it is expressed in terms of radiation components only, which share the same scaling, and is thus conserved after the full curvaton decay. 
Moreover, it is easy to check that, in the sudden decay approximation, $r_s$ reduces to $r_{\rm dec}$ given in Eq. (\ref{r_dec}), as $\bar\rho_{r \sigma, f}$ and $\bar\rho_{r \phi, f}$ can be replaced with $\bar\rho_{\sigma,{\rm dec}}$ and $\bar\rho_{r, {\rm dec}}$, respectively.

Interestingly,  the parameter $r_s$ can  also be directly related to the entropy  produced from the curvaton decay. Indeed, it can be written as
\begin{equation}\label{eq:entropy}
\begin{split}
r_s= 1 - \left( 1 + \frac{3}{4} q_s \right)^{-1}, \qquad
q_s
\equiv \left( \frac{ S_f}{S_i} \right)^{4/3} - 1,
\end{split}
\end{equation}
where $S_i$ and $S_f$
are respectively the initial  and final entropies  in a comoving volume\footnote{
The symbol $S$ (for the total entropy) should not be confused with $\mathcal{S}$ (for the isocurvature perturbation).
}.
Thus $r_s$ is more adapted to describe the curvature perturbations in a more general context. 
In Appendix \ref{sec:rs}, we show a derivation of the sudden decay formula
by using $r_s$ for the constant decay rate case, where we show that, for the case with a constant decay rate,
the final curvature perturbation $\zeta$ is given as the same form as Eq.~\eqref{eq:zeta1st} just by 
replacing $r_{\rm dec}$ with $r_s$.

\begin{figure}[htbp]
\begin{center}
\includegraphics[width=78mm]{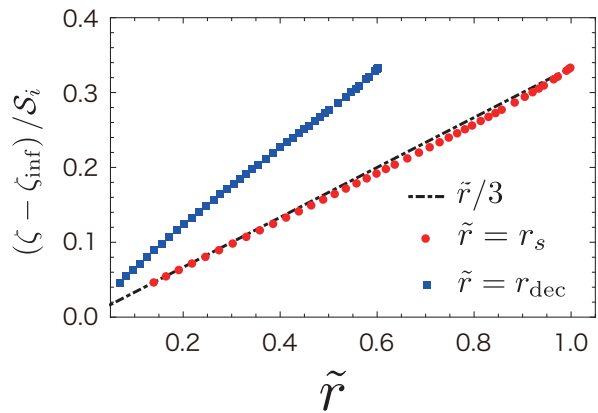}
\end{center}
\caption{Numerically evaluated curvature perturbation, $(\zeta-\zeta_{\rm inf})/{\mathcal S}_i$ in the curvaton scenario with constant decay rate. 
We take $m_\sigma = 10^{-16} M_{\rm P}$, $\Gamma  = 10^{-12} m_\sigma$ for the numerical calculation. 
$\tilde{r}$ is varied by changing $\bar{\sigma}_i$.   }
\label{fig:fit}
\end{figure}

In Fig.~\ref{fig:fit}, we plot of the curvature perturbation as a function of 
$r_{\rm dec}$ or $r_s$ for the case with a constant decay rate.
In this figure, black dot-dashed line shows the standard sudden decay analytic formula  $\tilde{r}/3$, the blue box 
and red circle show the numerical results as a function of numerically evaluated $r_{\rm dec}$ and $r_s$ given by Eq. (\ref{eq:entropy}), 
respectively.
As shown in this figure, if one uses $r_s$, the analytic formula $\zeta = (r_s/3){\cal S}_i$ can well describe the numerically obtained $\zeta$.
On the other hand, the use of $r_{\rm dec}$ does not give a good description, particularly, when $r_{\rm dec}$ is large.
(We should also note that both parameters give a good description of $\zeta$ when $\tilde{r} \ll {\cal O}(1)$.) 
Furthermore,  even when $r_s$ reaches unity, $r_{\rm dec}$ does not because of the existence of radiation produced by 
the curvaton. Therefore, we can see that the sudden decay formula with $r_s$ can describe $\zeta$ in this respect as well.

\subsection{Temperature-dependent decay rate}

Given some initial conditions for the energy densities and some specific function $\Gamma(T)$, one can easily  solve numerically the system of equations~(\ref{eq:evolvrho}) governing  the evolution of the curvaton and radiation energy densities.

\begin{figure}[htbp]
\begin{center}
\includegraphics[width=78mm]{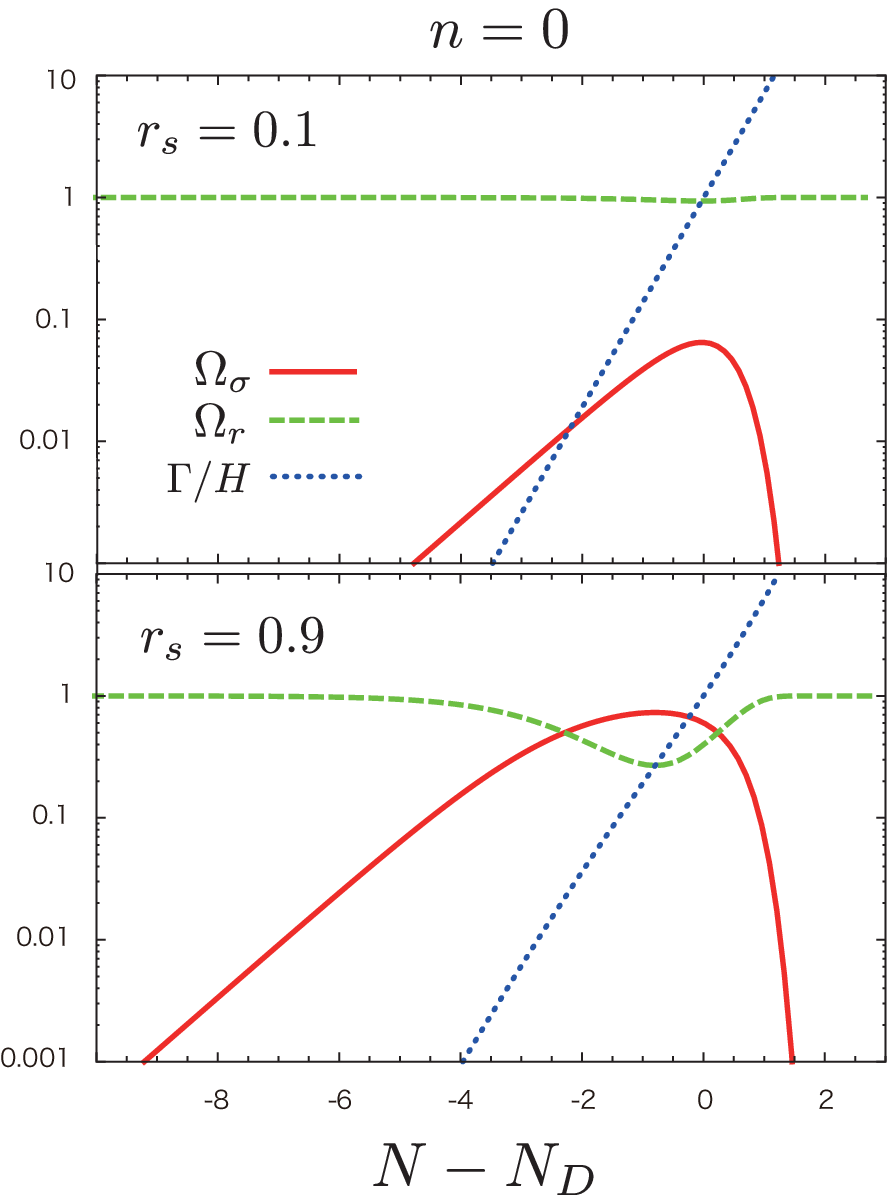}
\includegraphics[width=78mm]{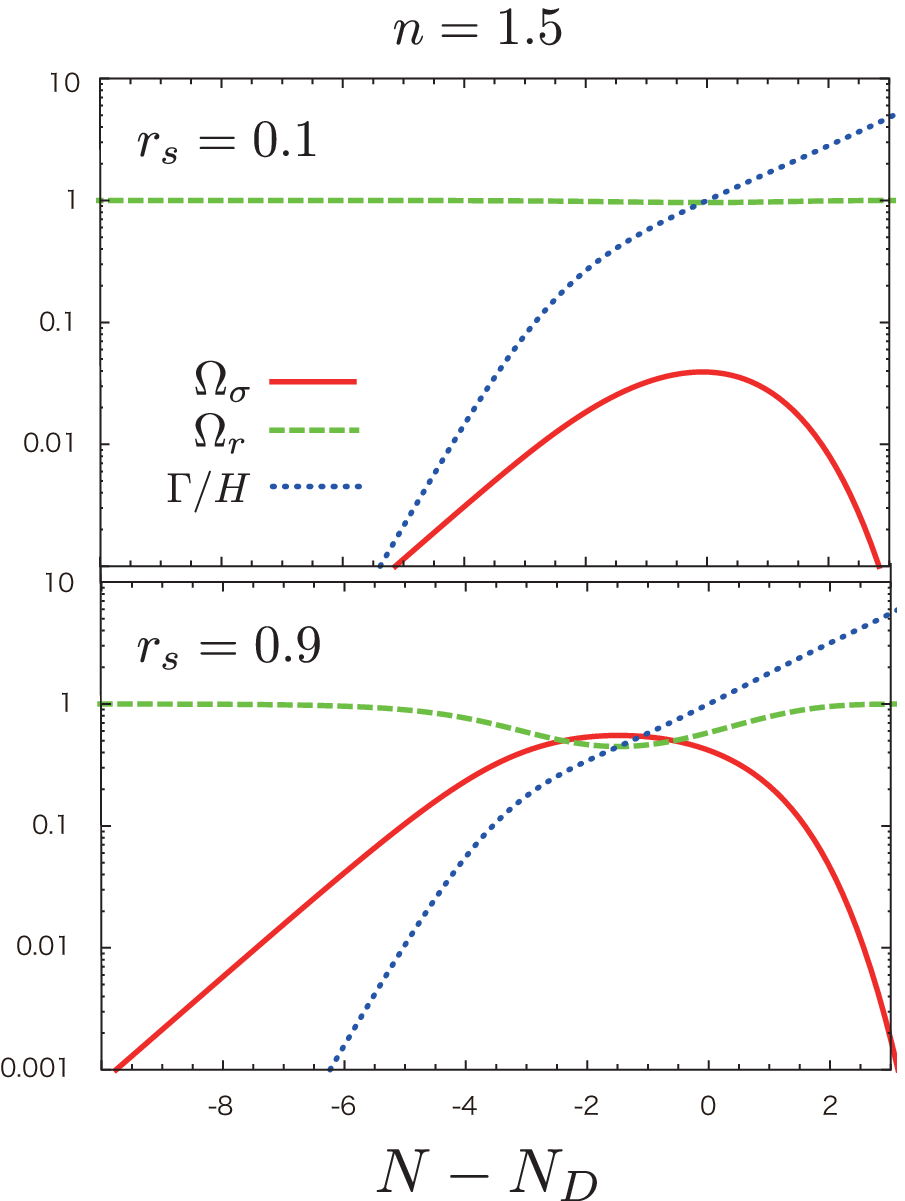}
\end{center}
\caption{Numerical results for the background evolution of $\Omega_{\sigma}$, $\Omega_r$ and $\Gamma / H$ for the cases of 
 $n = 0$ (left panels), $1.5$ (right panels) with $r_s= 0.1$ (top) and $0.9$ (bottom). 
The parameters used in the numerical calculation are $m_\sigma = 10^{-16} M_{\rm P}$ and $\Gamma_0  = 10^{-12} m_\sigma$.
Other parameters are set as $A = 1$ and $C=0$ for the case with $n=0$, 
$A = 10^{-5}$ and $C = 10^{-3}$ for the case with $n=1.5$.
The value of  $\bar{\sigma}_i$ is chosen such that 
it gives $r_s= 0.1$ and $0.9$ for each panel.
Here, we use the $e$-folding number $N$ as a time coordinate, and $N_D$ corresponds to the time when $\Gamma = H$.}
\label{fig:evolve}
\end{figure}

Here we consider two examples 
with different decay rates $\Gamma(T)$: a constant one, i.e. $n=0$ and a temperature-dependent one with e.g. $n=1.5$ in Eq. (\ref{eq:Gamma}).
The evolutions of  $\Omega_{\sigma}, \Omega_r$ and $\Gamma / H$ are shown  in Fig.~\ref{fig:evolve} as a function of the number of $e$-folds. 
The left and right panels correspond to the cases  $n = 0$ and  $n=1.5$, respectively. 
In these examples, we have adjusted the initial conditions $\bar{\sigma}_i$ so that the parameter $r_s$, given by Eq. (\ref{eq:entropy}), is given as
$r_s = 0.1$ (top panels) and $0.9$ (bottom panels). This means that the net fraction of radiation created by the curvaton decay  is the same in both cases, even if the temperature dependence of $\Gamma$ is different.

Let us now investigate the cosmological perturbations about these background solutions.
To calculate the perturbations in the curvaton scenario, one can use either the $\delta N$ formalism~\cite{Lyth:2004gb,Starobinsky:1986fxa, Salopek:1990jq, Sasaki:1995aw, Sasaki:1998ug} or the standard cosmological perturbation theory
with multiple fluids
(see, e.g., 
Refs.~\cite{Malik:2008im, Malik:2002jb,Matarrese:2003tk,Malik:2006pm,Gupta:2003jc}).

In the $\delta N$ formalism, the curvature perturbation $\zeta$ on a uniform total energy density hypersurface  can be evaluated as
\begin{equation}
\begin{split}
\zeta (t) = N (t; \bar{\sigma}_i + \delta \sigma_i) - N (t; \bar{\sigma}_i),
\end{split}
\end{equation}
where $N (t;\sigma_i)$ is the $e$-folding number 
measured between the initial time $t_i$ on a flat hypersurface and the final time $t_f$
on a uniform total energy density one.

The above expression is valid even at nonlinear order and can also be used to compute, in the case of local non-Gaussianity\footnote{
Throughout this paper, we  consider only the so-called local  non-Gaussianity, which is relevant in the curvaton scenario.},  the non-linearity parameter $f_{\rm NL}$ defined  by writing $\zeta$, up to  the second order, in the form
\begin{eqnarray}
\zeta = \zeta_{\rm G} + {3 \over 5} f_{\rm NL}\left( \zeta_{\rm G}^2 - \langle \zeta_{\rm G}^2\rangle \right),
\end{eqnarray}
where $\zeta_{\rm G}$ represents the linear part of $\zeta$ which obeys pure Gaussian statistics.
Using the Taylor expansion of $N (t;\sigma_i)$ and Wick's theorem (the fluctuations $\delta \sigma_i$ being treated as purely Gaussian), one  easily finds
\begin{eqnarray}
\label{f_NL}
f_{\rm NL} = {6 \over 5} {N_{\sigma\sigma} \over N_\sigma^2},
\end{eqnarray}
where $N_\sigma$ and $N_{\sigma\sigma}$  respectively denote $N_\sigma := \partial N (t; \bar{\sigma}_i) / \partial \bar{\sigma}_i$
and $N_{\sigma \sigma} := \partial^2 N (t;\bar{\sigma}_i) / \partial \bar{\sigma}_i^2$. Note that the expression (\ref{f_NL}) applies when the curvaton contribution in $\zeta$ is dominant over the inflaton contribution.

In order to cross-check our numerical results, we have also adopted the standard cosmological perturbation theory with multiple fluids and  solved numerically on superhorizon scales. In this case, the evolution equations depend on the curvature perturbation $\zeta$ and the isocurvature one ${\mathcal S}$, 
which can be expressed in terms of  fluctuations of each component, $\delta_a \equiv \delta \rho_a / \bar{\rho}_a$, as
\begin{eqnarray}
\zeta = \frac{ \bar{\rho}_{\sigma} \delta_{\sigma} +  \bar{\rho}_r \delta_r}{3  \bar{\rho}_{\sigma} + 4  \bar{\rho}_r}, \qquad {\mathcal S} = -3 \left( {\bar{H} \bar{\rho}_\sigma \over \bar{\dot{\rho}}_\sigma} \delta_\sigma -  {\bar{H} \bar{\rho}_r \over \bar{\dot{\rho}}_r} \delta_r \right).
\end{eqnarray}%
Following Ref.~\cite{Malik:2008im}, we have also numerically evaluated the evolution of $\zeta$ and ${\mathcal S}$
at second order in the perturbation theory. 
We have checked  that these two approaches lead to the same results for $\zeta$ and $f_{\rm NL}$.

\begin{figure}[htbp]
\begin{center}
\includegraphics[width=78mm]{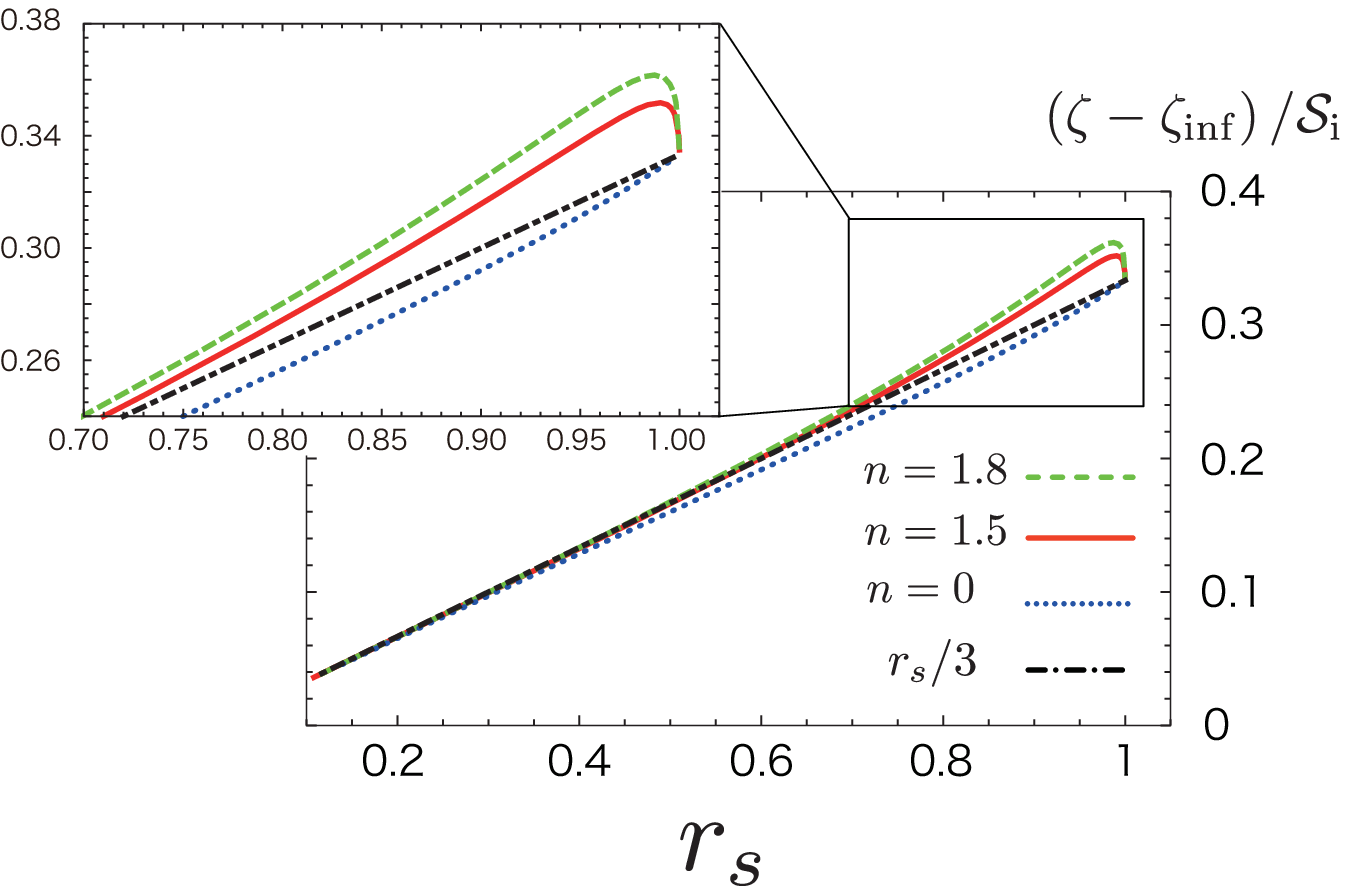}
\includegraphics[width=62mm]{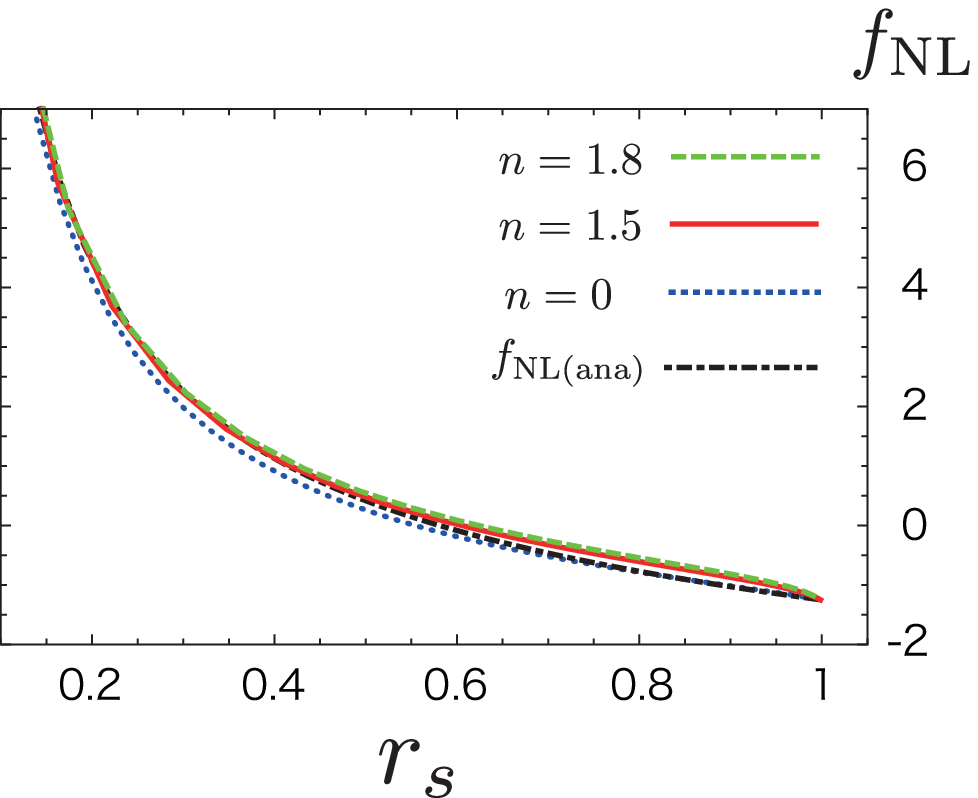}
\end{center}
\caption{The final curvature perturbation $\zeta$  (left panel) and the non-linearity parameter $f_{\rm NL}$ (right panel)
 are shown as a function of the parameter $r_s$.
We have plotted the cases with three different temperature dependences for $\Gamma (T)$, i.e., 
  $n = 0$ (blue dotted), $1.5$ (red solid) and $1.8$ (green long-dashed).
The  parameters $m_{\sigma}$, $\Gamma_0$, $A$ and $C$ are the same as in Fig.~\ref{fig:evolve} for $n=0$ and $n=1.5$.
For $n=1.8$ case, we have used the same parameters as those for $n=1.5$.
The value of  $r_s$ is varied by tuning $\bar{\sigma}_i$. 
For comparison, we have also plotted the analytic estimates given in (\ref{ana_estimate_1})  and (\ref{ana_estimate_2}) as black/dot-dashed lines.
}
\label{fig:zeta}
\end{figure}
In Fig.~\ref{fig:zeta}, we show the  final curvature perturbation $\zeta$ (left panel) and the non-linearity parameter $f_{NL}$ (right panel) 
in terms   of the  parameter $r_s$. We consider three different  decay rates with $n=0$ (blue dotted), $1.5$ (red solid) and $1.8$ (green long-dashed). 
For comparison, we  plot  the  analytic expression obtained in the standard 
curvaton scenario, i.e. \eqref{eq:zeta1st} with $\delta_\Gamma=0$, with the global transfer parameter $r_s$, which gives a much better estimate in the $n=0$ case, as discussed previously:
\begin{equation}
\label{ana_estimate_1}
\frac{\zeta-\zeta_{\rm inf}}{\mathcal{S}_i}  = \frac{r_s}{3}\,.
\end{equation}
We postpone the comparison with the full sudden decay expression, i.e. including $\delta_\Gamma$, to the next subsection.
Similarly, in the right panel, we  plot  the analytical expression $f_{\rm NL}$  (see, e.g., \cite{Sasaki:2006kq}), 
\begin{equation}
\label{ana_estimate_2}
 f_{\rm NL (ana)} = {5 \over 4 r_s} - {5 \over 3} - {5 \over 6}r_s\,,
\end{equation}
where, once again, we have replaced the sudden decay parameter $r_{\rm dec}$ by our new transfer parameter $r_s$.

As seen from the left panel, 
the analytic formula (\ref{ana_estimate_1}) can well describe the resultant curvature perturbations  for small values of $r_s$,
even in the case with temperature dependent decay rates.
The limit $r_s \ll 1$ corresponds to the situation where the curvaton is always subdominant at its decay.  This implies that the Hubble parameter is determined only by the temperature of  radiation from the inflaton. Since the decay is characterized by $H=\Gamma$, one expects the decay to occur at the same temperature $T_D$ in this limit
even for the time-dependent $\Gamma$ case. This effectively corresponds to the constant decay rate case, with  $\Gamma=\Gamma(T_D)$. This explains why all curves coincide in the limit $r_s\ll 1$. 
In the opposite limit, i.e.  $r_s=1$, one sees that the numerical result  approaches the simple analytic solution, $r_s / 3$, which does not take into account the temperature dependence of $\Gamma$.
This is  simply because at the time of curvaton decay the energy density of the curvaton field
dominates and it is just the same situation as the basic reheating picture of the inflaton
where it is known that no thermal effect affects the amplitude of the final curvature perturbation~\cite{ArmendarizPicon:2003pm,Weinberg:2004kr}.

For relatively large values of $r_s$,  one observes a deviation between the numerical result and the analytical estimate, which can reach about $5 \%$ in the $n=0$ case and becomes larger as $n$ increases. 
As discussed in the previous section,  the sudden decay expression Eq.~\eqref{eq:atdecay} would in fact suggest a strong enhancement, due to the $\delta_\Gamma$ term, for some parameter values. But, as expected,   such a large enhancement does not appear  in the numerical calculations, even if the behaviour of  $\zeta$ exhibits a mild enhancement. 
In the next subsection, we will investigate in more detail the discrepancy between the sudden decay approximation and the numerical computations.

Before closing this subsection, let us  briefly discuss the non-linearity parameter $f_{\rm NL}$ 
for the case with a temperature dependent $\Gamma$. We plot $f_{\rm NL}$ as a function of $r_s$ 
in the right panel of Fig.~\ref{fig:zeta}, in which 
one can see that  $f_{\rm NL}$ is hardly sensitive to  the temperature dependence, even for large values of $r_s$. 
Given the uncertainty of current and future cosmological experiments, typically  $\Delta f_{\rm NL} = \mathcal{O} (1)$\footnote{
We note that future observations of 21 cm fluctuations may probe $f_{\rm NL}$ more precisely \cite{Cooray:2006km,Chongchitnan:2012we}.
}, one concludes that one can safely neglect the impact of  the temperature-dependence of the decay rate on  the non-linearity of the curvature perturbation in the curvaton scenario. 

\subsection{Regime of validity of the sudden decay approximation}

\begin{figure}[htbp]
\begin{center}
\includegraphics[width=120mm]{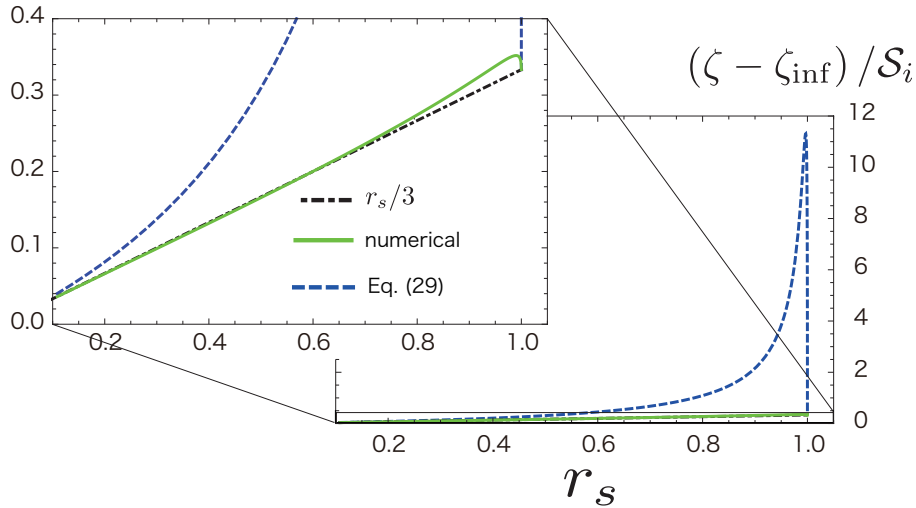}
\end{center}
\caption{The final curvature perturbation as a function of $r_s$ for $n=1.5$. 
A black dot-dashed line corresponds to the analytic formula given by Eq.~\eqref{ana_estimate_1}, 
a green solid line shows the result obtained by numerical calculation, and a blue dashed line corresponds to Eq.~(\ref{ana_estimate_3}).
The  parameters $m_{\sigma}$, $\Gamma_0$, $A$ and $C$ are the same as in Fig.~\ref{fig:zeta}.
Here, we have used an sudden decay expression for $\alpha$ which is given in Appendix \ref{sec:tdec}.
}
\label{fig:delta}
\end{figure}

Let us now concentrate on the differences between the sudden decay expressions obtained in the previous section and our numerical results.  As already mentioned,  the sudden decay analytical formula  (\ref{eq:zeta1st}), together with Eq.~(\ref{eq:dGS}), predicts a large enhancement of the curvature perturbation for some parameter values, which is not observed numerically. This is illustrated in Fig. \ref{fig:delta}, where we compare the numerical results with the analytical expression
\begin{equation}
\label{ana_estimate_3}
\begin{split}
\zeta = \zeta_{\text{inf}} + \frac{r_s}{3} \mathcal{S}_i - \frac{r_s}{6} \delta_{\Gamma},\qquad \delta_\Gamma = -  \alpha \, \frac{3 r_s}{3+r_s}
\left( \frac{3}{3+r_s}- \frac{\alpha}{2} \right)^{-1} {{\mathcal S_i} \over 3} \,,
\end{split}
\end{equation}
which corresponds to (\ref{eq:zeta1st}) and (\ref{eq:dGS}), with $r_{\rm dec}$ replaced by $r_s$.
In fact,
it turns out that the ``naive'' sudden decay formula Eq.~\eqref{ana_estimate_1} that neglects the  $\delta_\Gamma$ term, i.e. that does not take into account the fluctuations of $\Gamma$, provides a better approximation than the one with the temperature effect being included (i.e., $\delta_\Gamma \ne 0$), 
 for a wide range of values of $r_s$. However we should also note that,  as $r_s$ increases, particularly when $r_s\gtrsim 0.7$, 
 one observes that the numerical result gradually  deviates from Eq.~\eqref{ana_estimate_1}.

%

\begin{figure}[htbp]
\begin{center}
\includegraphics[width=78mm]{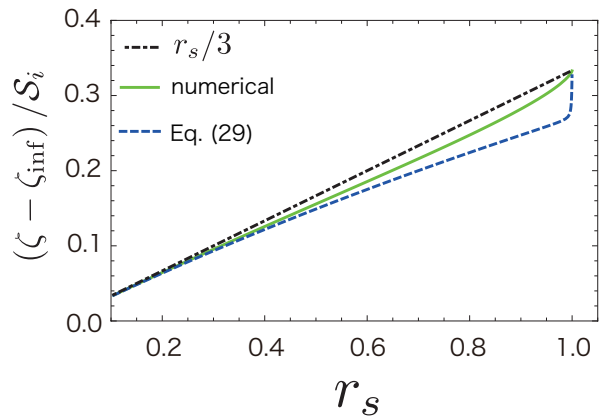}
\end{center}
\caption{The amplitude of the curvature perturbation as a function of $r_s$ for $n=-5$. 
The numerical result (green solid line) is compared to the analytic expression \eqref{ana_estimate_1}
(black dot-dashed line) and  to Eq.~(\ref{ana_estimate_3})  (blue dashed line). 
The parameters used in the numerical calculation are $m_\sigma = 10^{-16} M_{\rm Pl}$, $\Gamma_0 = 10^{-12} m_\sigma$, $A=10^{-5}$ and $C = 10^{-2}$, and
then $r_s$ corresponds one-to-one with $\bar{\sigma}_i$. 
}
\label{fig:negative}
\end{figure}
So far, we have considered only a temperature dependent decay rate with $n>0$.
Although not very realistic from a physical point of view, it is nevertheless instructive to study the case where the temperature dependence of $\Gamma$ is characterized by a negative power index $n < 0$, in order to compare the sudden decay expressions with the numerical computation in a wider range of situations. 
In Fig.~\ref{fig:negative}, we show the amplitude of the curvature perturbation as a function of $r_s$ for $n=-5$. We see that the numerical result is located between the analytical predictions (\ref{ana_estimate_1}) and (\ref{ana_estimate_3}). In contrast with the $n>0$ cases, the  expression (\ref{ana_estimate_3}) remains close to the numerical result and provides a better estimate of  the numerical result  for $r_s \lesssim 0.4$,
  even if the simple expression  (\ref{ana_estimate_1}) is more accurate for larger values of $r_s$.

\begin{figure}[htbp]
\begin{center}
\includegraphics[width=78mm]{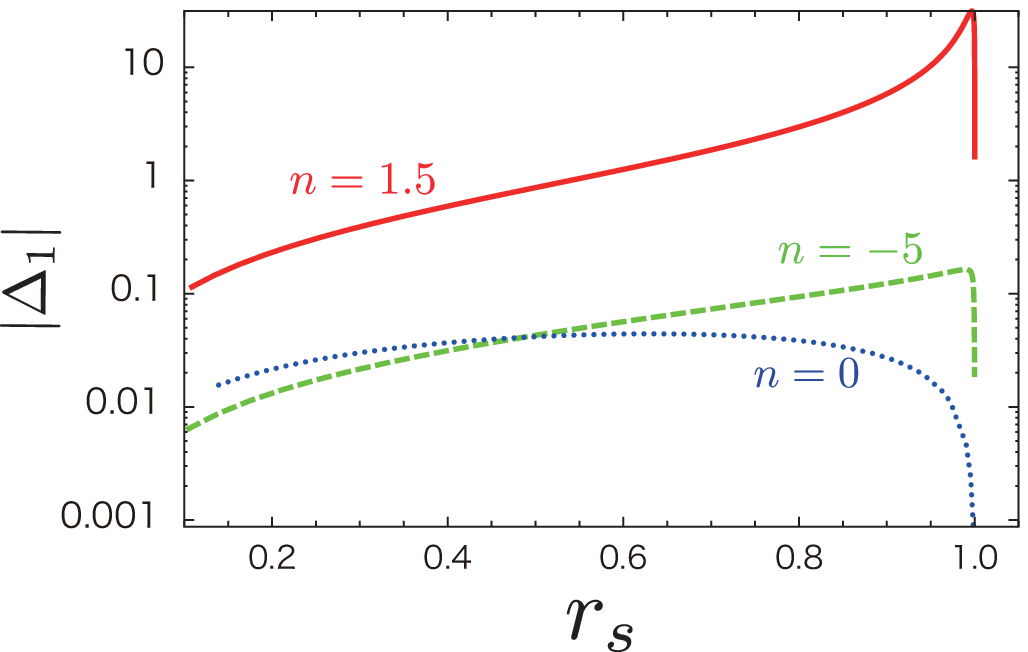}
\includegraphics[width=78mm]{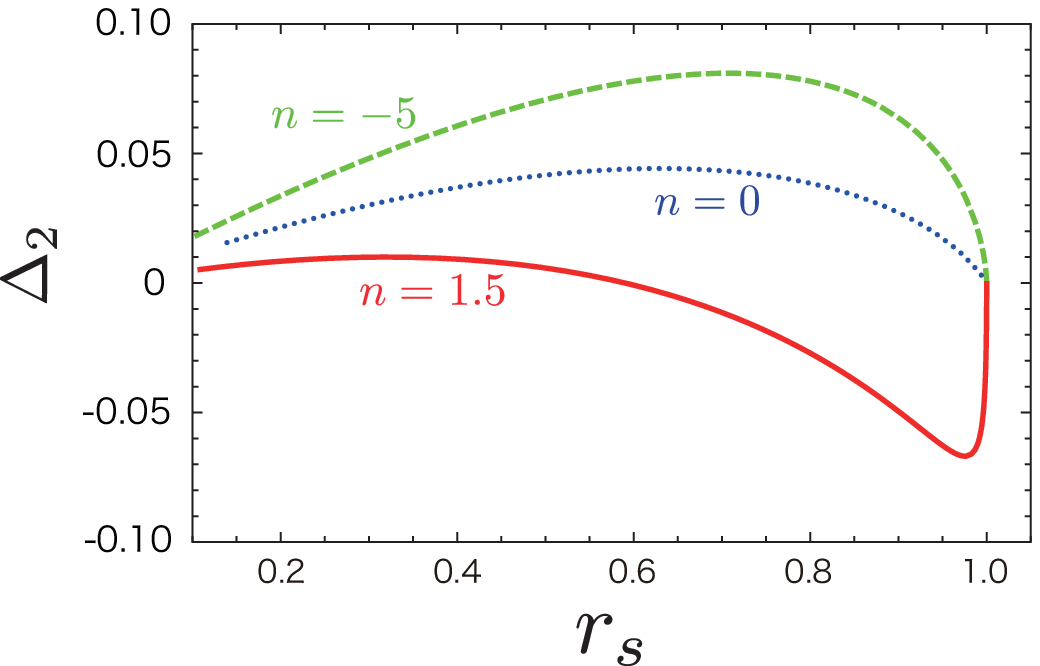}
\end{center}
\caption{ The relative difference between the numerical result and the analytic formulae for each case as a function of $r_s$.
The left panel shows the relative difference between the numerical result, $\zeta_{\rm num}$, and the analytical formula given by Eq. (\ref{ana_estimate_3}), and the right one shows the relative difference between $\zeta_{\rm num}$ and the simple  formula given by Eq. (\ref{ana_estimate_1}).
Notice that we plot the absolute value of $\Delta_1$ in logarithmic scale in the left panel and $\Delta_2$ in linear scale in the right panel.
}
\label{fig:reladif}
\end{figure}

To summarize the differences between the numerical result and the analytic expressions,
we plot the relative differences between these in Fig.~\ref{fig:reladif}.
The left panel shows the relative difference between the numerical result $\zeta_{\rm num}$, and 
the analytical formula Eq. (\ref{ana_estimate_3}), denoted $\zeta_{{\rm ana},1 }$ here, which is defined as
\begin{eqnarray}
\Delta_1 = {\zeta_{{\rm ana},1} \over \zeta_{\rm num}} - 1.
\end{eqnarray}
The right panel shows the relative difference between $\zeta_{\rm num}$ and the one given by 
the simple  formula Eq.~(\ref{ana_estimate_1}), denoted  $\zeta_{{\rm ana},2}$,  as
\begin{eqnarray}
\Delta_2 = {\zeta_{{\rm ana},2} \over \zeta_{\rm num}} - 1.
\end{eqnarray}
As seen from the figure, the differences $\Delta_1$ and $\Delta_2$ both decrease as $r_s$ approaches  $0$ or $1$.
However, it should be noticed that the deviation at $r_s \sim \mathcal{O}(0.1)$ is larger for $\Delta_1$ compared to $\Delta_2$. 
This shows that the simple sudden decay formula without $\delta_\Gamma$ gives a better approximation than that with $\delta_\Gamma$,
which means that the sudden decay formula including the effect of $\delta_\Gamma$ overestimate the amplitude of $\zeta$.

For a  deeper understanding  of the numerical results, we now  examine the system  of equations governing the evolution 
of the curvature  and isocurvature perturbations using the standard cosmological   
 perturbation theory with multiple fluids.  The evolution equations for $\zeta$ and ${\mathcal S}$ are 
 given by \cite{Malik:2008im,Malik:2002jb,Matarrese:2003tk,Gupta:2003jc,Malik:2006pm}
\begin{equation}
{d \zeta \over dN} = {\cal T}_{\zeta}\  {{\mathcal S} \over 3} , \qquad
{d  {\mathcal S} \over dN} = {\cal T}_{\mathcal S} \  {\mathcal S} ,
\label{eq:perturbation}
\end{equation}
with  the time-dependent coefficients  ${\cal T}_\zeta$ and ${\cal T}_{\mathcal S}$  
\begin{eqnarray}
{\cal T}_\zeta &\equiv& \left({3 - 2 g \over 3(1- g)}\right) \left({4 - {4-3g \over 1 - g} \Omega_\sigma \over 4 - \Omega_\sigma} \right)
\left( {3 \Omega_\sigma \over 4 - \Omega_\sigma} \right), \cr\cr
{\cal T}_{\mathcal S}  &\equiv& - {g \over 2(1-g)} 
{4 (1-g) - (4-3g) \Omega_\sigma \over 3 - 2 g}  \cr\cr
&&\times \Biggl[ 1 
+ \left({ 3 - 2g\over 4 (1-g) - (4-3g) \Omega_\sigma} \right)^2 \Omega_\sigma (2 - \Omega_\sigma)
- {\alpha \over 2(1 - \Omega_\sigma)}\Biggr] ,
\end{eqnarray}
where $g(N) \equiv \Gamma/(\Gamma + H)<1$.
We can formally integrate the above equation, which leads to  an expression for  $\zeta$ of the form
\begin{equation}
\zeta = \zeta_{\rm inf} + \int dN {\mathcal F}(N)\,  {{\mathcal S}_i \over 3}, 
\end{equation}
where
\begin{equation}
\label{eq:F}
{\mathcal F}(N) \equiv {\cal T}_\zeta(N) \exp \left[ \int^N dN' \, {\cal T}_{\mathcal S} (N') \right]\,.
\end{equation}
%
\begin{figure}[htbp]
\begin{center}
\includegraphics[width=125mm]{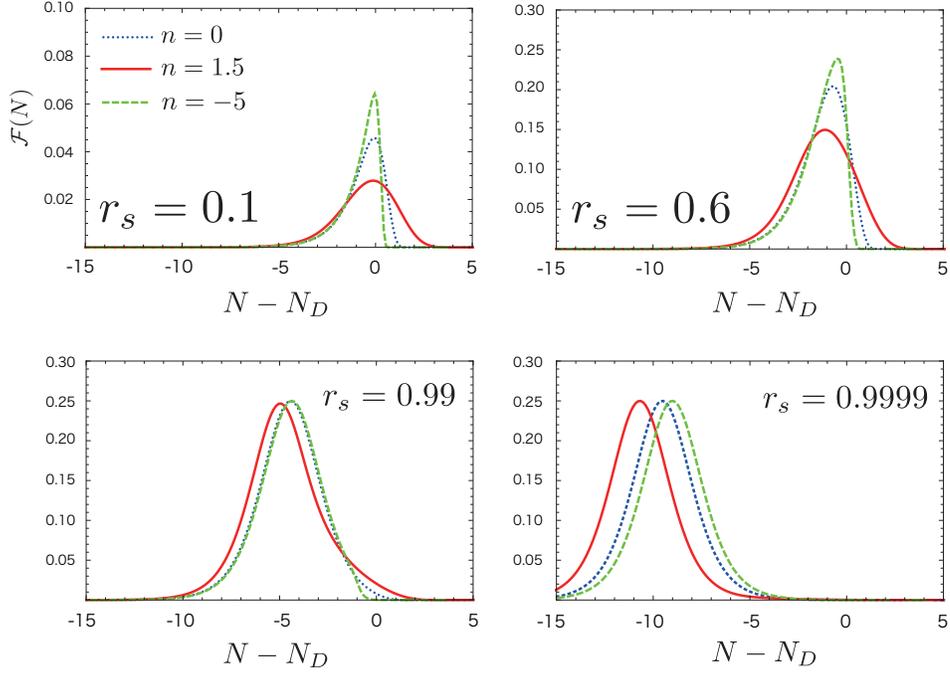}
\end{center}
\caption{Plots of ${\mathcal F}(N)$ as a function of the $e$-folding number $N - N_D$ with $N_D$ corresponding to
the decay time which is defined as the epoch satisfying $H=\Gamma$  for the cases with $n=0$ (blue dotted), $1.5$ (red solid) and $-5$ (green dashed). 
We show the plots for the several values of  $r_s$:  $r_s = 0.1$ (top left), $0.6$ (top right), $ 0.99$ (bottom left) and $ 0.9999$ (bottom right).
}
\label{fig:source}
\end{figure}
%

%
\begin{figure}[htbp]
\begin{center}
\includegraphics[width=78mm]{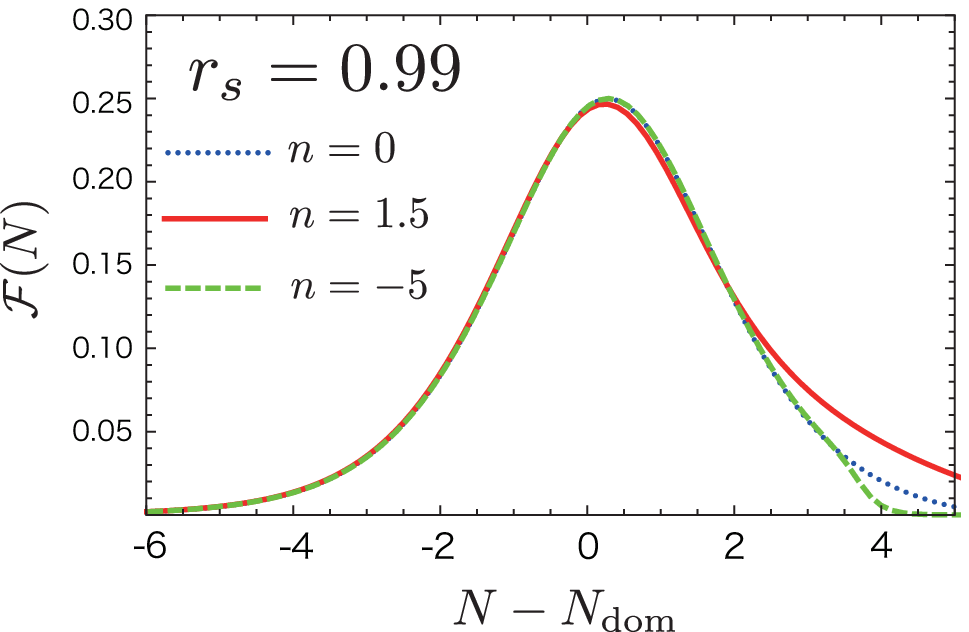}
\includegraphics[width=78mm]{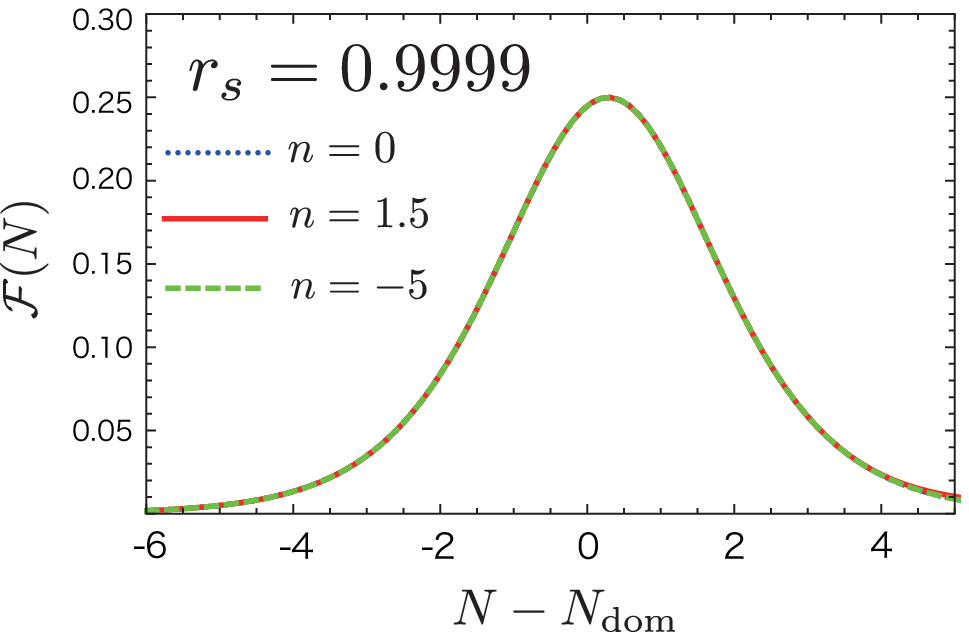}
\end{center}
\caption{Plots of ${\mathcal F}(N)$ as a function of the $e$-folding number $N - N_{\rm dom}$ with $N_{\rm com}$ corresponding to
the time when the curvaton begins to dominate the energy density of the Universe. More specifically, 
$N_{\rm dom}$ is defined as the time at which $\Omega_\sigma = 1/2$ is realized. Here we show the cases with $r_s = 0.99$ (left) 
and $0.9999$ (right).}
\label{fig:source_Ndom}
\end{figure}
%

In Fig.~\ref{fig:source}, 
we show the transfer function ${\mathcal F}(N)$ defined in Eq.~(\ref{eq:F}) as a function of the $e$-folding number measured from the decay time $N_D$, corresponding to the instant at which $H=\Gamma$.
First let us focus on the cases with relatively small values of $r_s$ shown in the top panels of Fig.~\ref{fig:source}.
For such values of  $r_s$,  the energy density of the Universe is still dominated, at the curvaton decay, by the radiation component produced from the inflaton.
From the figure, we  see that 
the transfer function ${\mathcal F}(N)$ has a peak around the decay time $N= N_D$.
We  also find that  the height of the peak becomes lower and
the ``width" of the function ${\mathcal F}(N)$ becomes broader as
the value of $n$ increases.

Next, let us consider the cases where $r_s$ is close to unity. 
In Fig.~\ref{fig:source}, we show  the cases with  $r_s = 0.99$ (bottom left) and $0.9999$ (bottom right).
For such cases,  the curvaton energy density starts to dominate the Universe long before the curvaton decay.
Interestingly, the peak position of the transfer function corresponds to 
the time when the curvaton begins to dominate, not at the decay time. 
Furthermore, the peak position shifts to smaller values of $N$ as $n$ increases, which 
comes from the fact that the curvaton begins to dominate the Universe earlier when $n$ is larger. 
To illustrate that the peak position indeed corresponds to the time of the curvaton domination,  we 
 plot in Fig.~\ref{fig:source_Ndom} the transfer function ${\cal F}(N)$ as a function of $N - N_{\rm dom}$ where $N_{\rm dom}$ is defined 
as the time when $\Omega_\sigma =1/2$.  From the figure, we can clearly see that 
the position of the peak almost corresponds to the domination time $N_{\rm dom}$ and this 
tendency does not depend on the functional form of $\Gamma(T)$. 
Furthermore, by comparing the plots for the cases with $r_s = 0.99$ and $0.9999$,
one can also notice that, as $r_s$ approaches  unity, 
the transfer function becomes identical regardless of the value of $n$. 
And then, as $r_s$ decreases from 1, the tail of the transfer function is broader for larger $n$.

%
\begin{figure}[htbp]
\begin{center}
\includegraphics[width=78mm]{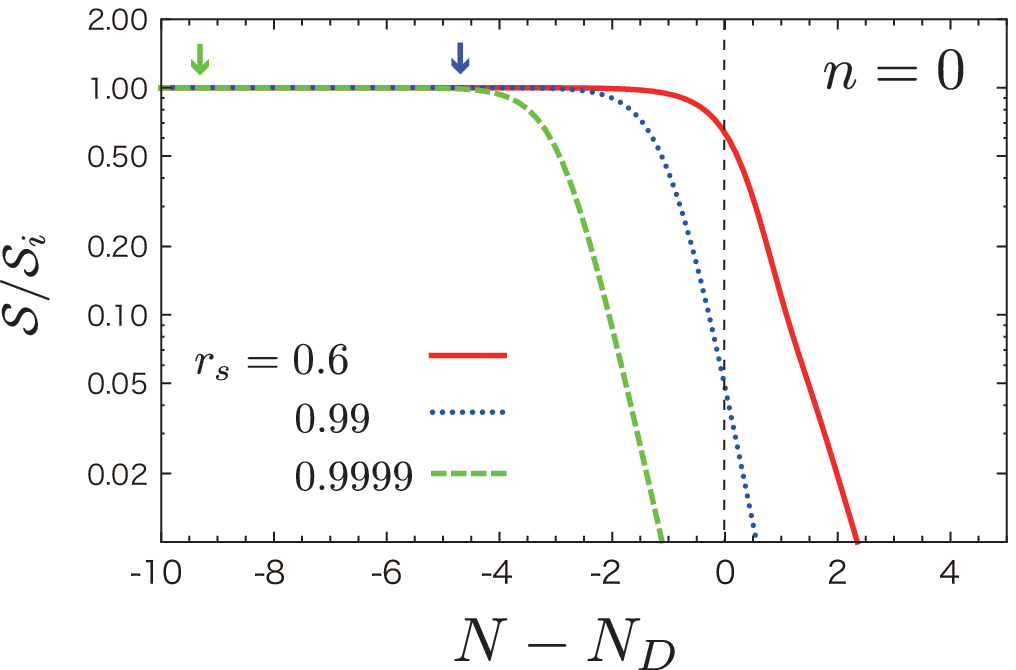}
\includegraphics[width=78mm]{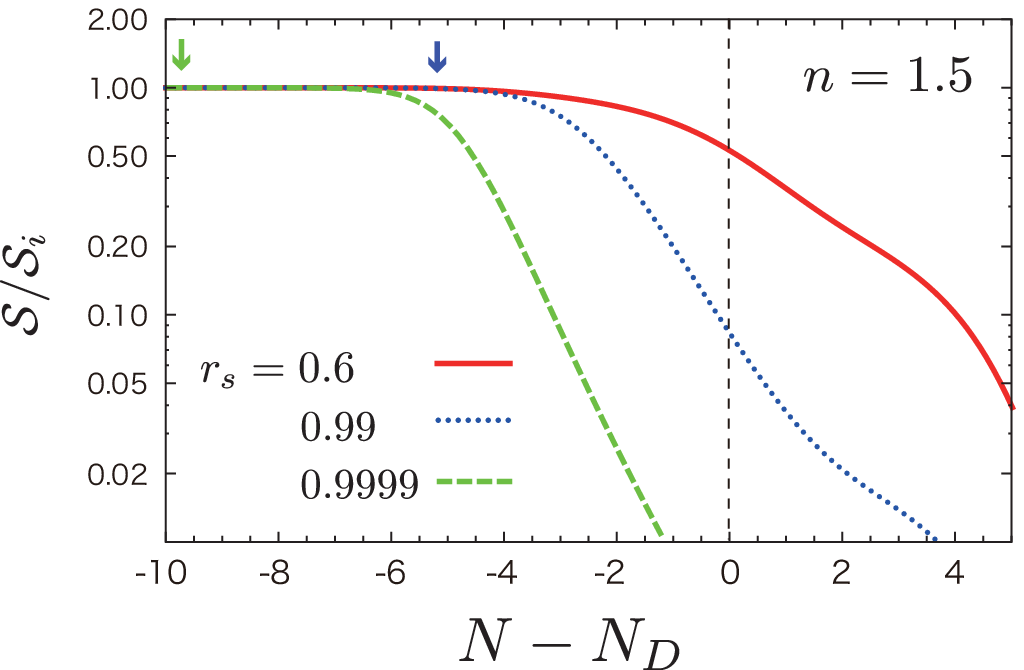}
\includegraphics[width=78mm]{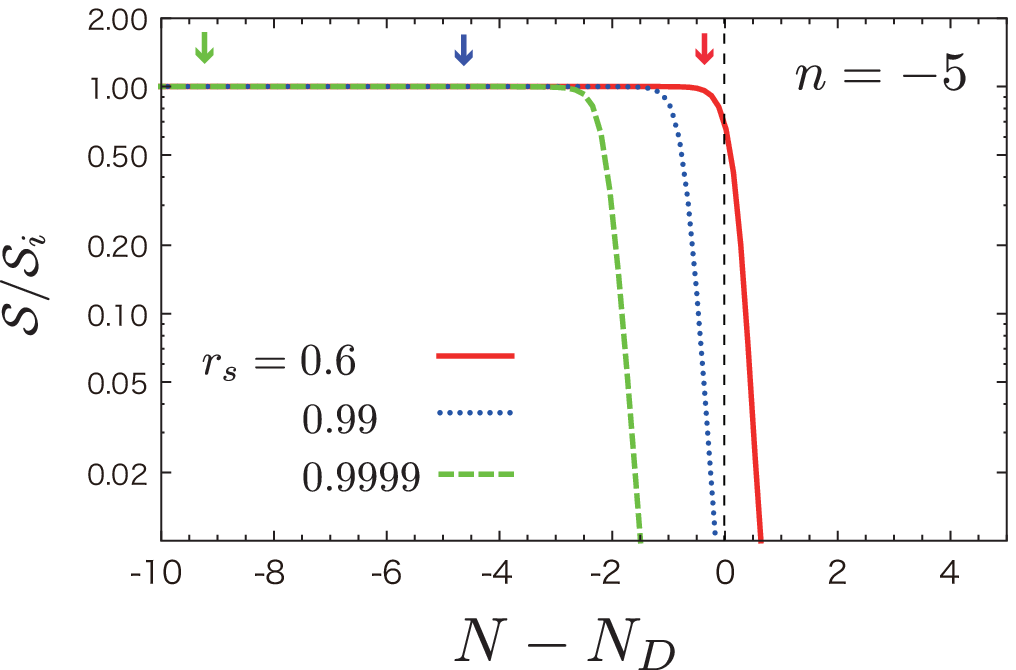}
\end{center}
\caption{Evolution of the isocurvature perturbation as a function of the $e$-folding number $N - N_D$.
The red solid line is for $r_s = 0.6$, where the curvaton energy density never dominates the Universe when $n=0$ and $1.5$. 
The blue dotted line and green dashed line are respectively for $r_s = 0.99$ and $r_s = 0.9999$, and each down-pointing arrow
represents the time when the curvaton energy density starts to dominate the Universe, which is defined as the time when $\Omega_\sigma = 1/2$
holds.
}
\label{fig:evo_iso}
\end{figure}
%
 
 For completeness, we also plot in Fig.~\ref{fig:evo_iso} the evolution of the isocurvature perturbation for several fixed values of $r_s$ and $n$.
As one can see in the figure, the amplitude of the isocurvature perturbation decreases during the curvaton decay phase, more or less early depending on the value of  $r_s$ and more or less rapidly depending on the value of $n$.  
In the transfer function ${\mathcal F}(N)$,  the evolution of the isocurvature perturbation, expressed by $\exp [\int dN' {\mathcal T}_S(N')]$, is combined 
with the function ${\mathcal T}_\zeta(N)$, which represents the transfer of the isocurvature into the adiabatic perturbations and also peaks at the decay since it is basically proportional to $\Omega_\sigma$.
Hence, 
the damping behavior of the isocurvature perturbation in Fig.~\ref{fig:evo_iso} helps to understand the ``shape" of the transfer function 
depicted in Fig.~\ref{fig:source}.
When the isocurvature perturbation begins to decrease earlier, 
the peak of the transfer function  ${\cal F}(N)$ also shifts to smaller value of $N$,
which can be noticed by comparing with the corresponding plot in Fig.~\ref{fig:source}.
Concerning the width of  ${\cal F}(N)$,
it tends to become broader when the isocurvature perturbation decreases slowly.
For example, looking at the case with $r_s=0.6$, one sees that ${\cal S}$ decreases more slowly as $n$ increases. 
This can be compared with the behavior of  the width of ${\cal F}(N)$.

\section{Conclusions}
\label{sec:conc}

In this work, we have explored how much  the thermal effects in the curvaton scenario could affect the primordial curvature perturbation. In  particular, we have  considered the temperature dependence of the curvaton dissipation rate, assuming  a simple  form for the function $\Gamma (T)$. 
We have first derived  analytical expressions for the final curvature perturbation in the sudden decay approximation.
We have also studied numerically the system consisting of the curvaton and radiation fluids and computed the evolution of both the background and the perturbations.  

We have found that the amplitude of  the final curvature perturbation $\zeta$ can be modified by as much as about ten percent in comparison with  the standard result with a constant $\Gamma$.
By contrast,  there is no significant effect  on  the non-linearity parameter $f_{NL}$, at least at the level of the latest precision reported by Planck.

Another important result of this work is the introduction of  a new definition of the transfer coefficient that relates the initial isocurvature perturbation (due to the curvaton) and the final curvature perturbation. In contrast with the usual definition given in the context of the sudden decay approximation, our  transfer parameter $r_s$ is defined globally and gives a very good fit to the numerical result, especially  in the 
  case of a constant decay rate.
For a temperature-dependent decay rate, the situation turns out to be  more complicated as one can derive within the sudden decay approximation, an a priori more refined expression, which takes into account the fluctuations of $\Gamma$ on the decay hypersurface. However, this expression leads to an artificially large enhancement due to an accidental degeneracy in the position of the decay hypersurface. This is confirmed by our numerical investigation, which shows that this refined expression is often a bad approximation.

As a final remark, we note that the thermal effects can modify not only the curvaton decay rate $\Gamma (T)$, but also  the curvaton potential itself, which has  important consequences for the curvaton dynamics.
For example, if the thermally generated effective mass of the curvaton dominates the potential,
the curvaton starts to oscillate earlier and the curvaton equation of state is modified~\cite{Enqvist:2013gwf}. It would also be interesting to investigate the consequences of such thermal effects on  the primordial curvature perturbation and may be pursued in the future.

\section*{Acknowledgments}
SY thanks Masahiro Kawasaki for useful comments.
This work of NK, TT~(T.Takesako) and SY was supported in part by JSPS Research Fellowships for Young Scientists. 
The work of TT~(T.Takahashi) is supported in part by Grant-in-Aid for Scientific Research 23740195
 from the Ministry of Education, Culture, Sports, Science and Technology in Japan.
DL was partly supported by the ANR (Agence Nationale de la Recherche) grant ``STR-COSMO" ANR-09-BLAN-0157-01. We would also like to thank the Yukawa Institute for Theoretical Physics at Kyoto University: discussions during the YITP workshop YITP-X-13-03 ``APC-YITP collaboration: mini-workshop on gravitation and cosmology" was useful to complete this work. 

\appendix
\section{Sudden decay formula with $r_s$}
\label{sec:rs}

Here, we show the derivation of the sudden decay formula by using the new definition $r_s$
for the constant decay rate case.
Let us consider that
 the curvaton energy density is totally converted into radiation just after the decay, so that
\begin{equation}\label{eq:sdc}
\begin{split}
\bar \rho_{\sigma}~\mathrm{e}^{3 (\zeta_{\sigma, i} - \delta N_D)} = \bar \rho_{r \sigma}~\mathrm{e}^{4 (\zeta_{r \sigma} - \delta N_D)},
\end{split}
\end{equation}
where  the subscript ``$r\sigma$" denotes the radiation produced by the curvaton decay (by contrast with the radiation already present before the decay).
Since $\bar\rho_{\sigma} = \bar \rho_{r \sigma}$ at $H = \Gamma$, we obtain the simple relation 
\begin{equation}\label{eq:curvatonsec}
\begin{split}
\delta N_D = 4 \zeta_{r \sigma} - 3 \zeta_{\sigma, i}.
\end{split}
\end{equation}
Denoting the radiation component  resulting from the inflaton decay by the subscript ``$r\phi$", 
the curvature perturbation $\zeta$ on a uniform total energy density hypersurface, at some time $t_f$ after the complete decay of the curvaton, is given  by the relation
\begin{equation}\label{eq:frelation}
\frac{ \bar\rho_{r \sigma, f}}{ \bar\rho_{r \sigma, f} +  \bar\rho_{r \phi, f}} \mathrm{e}^{4 (\zeta_{r \sigma} - \zeta)}
+ \frac{ \bar\rho_{r \phi, f}}{ \bar\rho_{r \sigma, f} +  \bar\rho_{r \phi, f}} \mathrm{e}^{4 (\zeta_{r \phi} - \zeta)}=1\,.
\end{equation}
At linear order, this yields
\begin{equation}\label{eq:frelationlinear}
\begin{split}
\zeta (t_f)
&= \frac{ \bar\rho_{r \sigma, f}}{ \bar\rho_{r \sigma, f} + \bar \rho_{r \phi, f}} \zeta_{r \sigma} 
+ \frac{ \bar\rho_{r \phi, f}}{ \bar\rho_{r \sigma, f} +  \bar\rho_{r \phi, f}} \zeta_{\text{inf}},
\end{split}
\end{equation}
where we have introduced the notation $ \zeta_{\text{inf}} \equiv  \zeta_{r\phi}$.
The relation between the total energy density and the decay rate of the curvaton
at the decay hypersurface, expressed in (\ref{rho_total_before}), can also be written as 
\begin{equation}\label{eq:justafter}
\begin{split}
\bar\rho_{\text{tot}}\,\mathrm{e}^{4 (\zeta - \delta N_D)}=3 M_{\text{P}}^2 \Gamma^2\,,
\end{split}
\end{equation}
and, at the linear order, we have
\begin{equation}
\zeta (t_f) = \delta N_D~.
\label{eq:zetadeltaN}
 \end{equation}
From Eqs.~(\ref{eq:curvatonsec}), (\ref{eq:frelationlinear}) and (\ref{eq:zetadeltaN}),
we obtain
\begin{equation}
\zeta (t_f) = \zeta_{\rm inf} + {r_s \over 3} {\mathcal S}_i.
\end{equation}

\section{Sudden decay approximation with $\Gamma=\Gamma(T)$}
\label{sec:tdec}

For the decay rate given by Eq. (\ref{eq:Gamma}),
 one can  write  $\alpha$ in terms of $r_{\rm dec}$ as
\begin{eqnarray}
\alpha = n \frac{\Gamma (T) - \Gamma_0 A}{\Gamma (T)} \frac{ 1 }{ 1 + C \left( T_{\rm dec} / m_\sigma \right)^n },
\end{eqnarray}
with
\begin{eqnarray}
\frac{T_{\rm dec} }{ m_\sigma } =\left(\frac{5}{72 \pi^2 g_*(T_{\rm dec})}\right)^{1/4} \left( \frac{m_\sigma}{M_{\rm P}}\right)^{-1/2} \left( \frac{\bar{\sigma}_i}{M_{\rm P}}\right)^2  {3(1-r_{\rm dec}) \over 4 r_{\rm dec}},
\label{eq:tdec}
\end{eqnarray}
where $g_*(T_{\rm dec})$ is the relativistic degrees of freedom at the decay time. 
By employing the sudden decay approximation, we show the derivation of the expression for $T_{\rm dec} / m_\sigma$ given by Eq. (\ref{eq:tdec}) as follows.
Here we define the cosmic temperature, $T$, from the energy density of the radiation and we have
\begin{eqnarray}
{T_{\rm dec} \over m_\sigma}
&=& {1 \over m_\sigma } \left( {30 \over \pi^2 g_* (T_{\rm dec})} \bar{\rho}_{r,{\rm dec}} \right)^{1/4},  \cr\cr
&=&   {1 \over m_\sigma } \left( {30 \over \pi^2 g_* (T_{\rm dec})} \bar{\rho}_{r,i} \left( {a_i \over a_{\rm dec}} \right)^{4}\right)^{1/4}.
\end{eqnarray}
Then, the scale factor can be related with the density parameter $\Omega_\sigma$ as
\begin{eqnarray}
\Omega_{\sigma,{\rm dec}} = {\bar{\rho}_{\sigma, i} \left( a_i / a_{\rm dec} \right)^3 \over 
\bar{\rho}_{\sigma, i} \left( a_i/ a_{\rm dec} \right)^3 + \bar{\rho}_{r,i} \left( a_i / a_{\rm dec} \right)^4} ,
\end{eqnarray}
and then we have
\begin{eqnarray}
{a_i \over a_{\rm dec}} =  \left( {\bar{\rho}_{\sigma , i} \over \bar{\rho}_{r,i}} \right) {1 - \Omega_{\sigma, {\rm dec}} \over \Omega_{\sigma, {\rm dec}}}.
\end{eqnarray}
Since at the initial time ($H_i = m_\sigma$) the Universe is dominated by the radiation energy density,
we have $\bar{\rho}_{r,i} \simeq 3 H_i^2 M_{\rm P}^2 = 3 m_\sigma^2 M_{\rm P}^2 $ and $\bar{\rho}_{\sigma, i} = {1 \over 2} m_\sigma^2 \bar{\sigma}_i^2$
and then
\begin{eqnarray}
{T_{\rm dec} \over m_\sigma} &=& {1 \over m_\sigma} \left( {30 \over \pi^2 g_*(T_{\rm dec})} {\bar{\rho}_{\sigma,i}^4 \over \bar{\rho}_{r,i}^3}\right)^{1/4}
 {1 - \Omega_{\sigma, {\rm dec}} \over \Omega_{\sigma, {\rm dec}}} \cr\cr
 &=&  \left( {5 \over 72 \pi^2 g_*(T_{\rm dec})} \right)^{1/4}
\left({m_\sigma \over M_{\rm P}}\right)^{-1/2} \left( {\bar{\sigma}_i \over M_{\rm P}} \right)^2  {1 - \Omega_{\sigma, {\rm dec}} \over \Omega_{\sigma, {\rm dec}}} \cr\cr
&=& \left( {5 \over 72 \pi^2 g_*(T_{\rm dec})} \right)^{1/4}
\left({m_\sigma \over M_{\rm P}}\right)^{-1/2} \left( {\bar{\sigma}_i \over M_{\rm P}} \right)^2 {3 (1 - r_{\rm dec}) \over 4 r_{\rm dec}}.
\end{eqnarray}

{}

\end{document}